\documentclass[aps,prb,reprint,superscriptaddress,nofootinbib,floatfix]{revtex4-2}
\usepackage{graphicx}
\usepackage{amsmath,amssymb}
\usepackage[colorlinks=true,allcolors=blue]{hyperref}

\begin{document}

\title{Superexchange coupling between a dilute Gd$^{3+}$ probe and orphan spins in a valence-bond glass}

\author{Patrick Colgan}
\author{Artur Solodovnyk}
\author{Jeremy P. Carlo}
\author{Scott Dietrich}
\email{scott.dietrich@villanova.edu}
\affiliation{Department of Physics, Villanova University, Villanova, Pennsylvania 19085, USA}

\date{\today}

\begin{abstract}
The valence-bond-glass state of the $4d^1$ double perovskite Ba$_2$YMoO$_6$ combines Mo$^{5+}$ singlets with residual paramagnetic ``orphan'' spins whose coupling is hard to reach with bulk probes. We probe them by electron paramagnetic resonance of a dilute Gd$^{3+}$ ion on the Y$^{3+}$ site via broadband coplanar-waveguide transmission at 5~K. The isostructural nonmagnetic Ba$_2$YNbO$_6$, doped with the same Gd$^{3+}$ probe, is an equivalent-lattice control. The Gd$^{3+}$ linewidth nearly doubles in the magnetic host. Dipolar coupling to the sparse orphans is nearly an order of magnitude too weak; the excess is instead reproduced by a transferred $f$--$d$ superexchange of $J \approx 0.2$~K, so the width is a static measure of orphan-spin coupling. The half-field ($\Delta m_S = \pm 2$) satellite collapses in the magnetic host, a dynamical dephasing channel, while the unchanged $g$-factor marks a zero-mean random internal field. Together the two channels probe the coupling and dynamics of valence-bond-glass orphan spins.
\end{abstract}


\maketitle

\section{Introduction}

Geometrically frustrated and quantum-disordered magnets, including spin liquids, spin ices, and valence-bond glasses, realize ground states that evade conventional long-range order. Characterizing them requires probes sensitive to local moments and their fluctuations, rather than to bulk-averaged order parameters~\cite{balents2010,broholm2020,wen2019}. Local magnetic resonance techniques such as nuclear magnetic resonance (NMR), muon spin relaxation ($\mu$SR), and electron paramagnetic resonance (EPR) complement thermodynamic and neutron-scattering measurements because they report the distribution of internal fields and its dynamics at the atomic scale~\cite{carlo2011}. One approach is to introduce a dilute, well-characterized probe spin whose resonance is set by the magnetic environment of the host. This converts a measurement of the host into a measurement of a single, calibrated line.

Rock-salt ordered double perovskites A$_2$BB$'$O$_6$~\cite{vasala2015,anderson1993} are well suited to this approach. The B$'$-site cations form a face-centered-cubic sublattice that is geometrically frustrated for antiferromagnetic B$'$-B$'$ interactions. The host magnetism can be switched on or off by chemical substitution on the B$'$ site while the lattice is left essentially unchanged. Ba$_2$YMoO$_6$, built on $4d^1$ Mo$^{5+}$, is a spin-orbit-entangled frustrated magnet~\cite{chen2010,witczak2014} that fails to order down to the lowest temperatures despite a large antiferromagnetic Curie--Weiss scale ($\theta_{CW} \approx -160$~K to $-220$~K). It instead condenses into a gapped, nonmagnetic singlet (valence-bond-glass) ground state, as established by magnetic susceptibility, heat capacity, NMR, neutron scattering, and $\mu$SR~\cite{devries2010,aharen2010,carlo2011}. Its nonmagnetic analogue Ba$_2$YNbO$_6$, built on $4d^0$ Nb$^{5+}$, is isostructural but carries no B$'$-site moment. The Nb/Mo pair therefore provides a magnetic versus nonmagnetic contrast at fixed crystal structure. This is the control needed to isolate a purely magnetic effect on a probe spin.

Gd$^{3+}$ ($4f^7$, $^8S_{7/2}$) is well suited for this purpose. Its half-filled shell gives a spin-only $S = 7/2$ moment with quenched orbital angular momentum ($L = 0$), so it shows only small zero-field splitting and a $g$-factor close to the free-electron value. Its resonance field and linewidth therefore report on the local magnetic field and its distribution rather than single-ion anisotropy~\cite{abragam1970}. Substituted dilutely onto the Y$^{3+}$ site of Ba$_2$(Gd$_x$Y$_{1-x}$)B$'$O$_6$, Gd$^{3+}$ occupies a well-defined site within the host B-sublattice and senses the internal fields produced by the host moments.

We measure the Gd$^{3+}$ resonance using broadband coplanar-waveguide (CPW) transmission spectroscopy rather than conventional cavity EPR. A CPW records the microwave transmission $S_{21}$ continuously over a wide frequency band as a function of applied field. This maps the resonance across many frequencies in a single experiment, requires no resonant cavity, and accommodates small polycrystalline samples in a planar geometry~\cite{kalarickal2006,bilzer2007,eckel2025}. Because the transmitted signal mixes the real and imaginary parts of the dynamic susceptibility, each resonance appears as a Dysonian (mixed absorption and dispersion) lineshape~\cite{dyson1955}. Over our frequency range every Gd$^{3+}$ line is accompanied by a weaker half-field satellite, and we analyze the two together.

In this work we use the electron paramagnetic resonance of a dilute Gd$^{3+}$ probe to perform a local spectroscopy of the orphan spins of the Ba$_2$YMoO$_6$ valence-bond glass, comparing the isostructural magnetic (Mo) and nonmagnetic (Nb) hosts at two Gd concentrations (2 and 5 atomic \%) and base temperature (5~K), with the field axis calibrated against a DPPH (2,2-diphenyl-1-picrylhydrazyl) standard. The Gd$^{3+}$ $g$-factor is $1.99(2)$ and independent of host and of Gd concentration, as expected for a local $S$-state ion, so the host effect appears in the linewidth rather than the resonance position. The main-line width nearly doubles in the magnetic host ($\approx 140$~mT versus $\approx 76$~mT), and its independence of Gd concentration shows that each probe reports the host coupling individually. Our central result is that this excess is not dipolar: with most Mo locked into singlets, the sparse orphan spins produce a dipolar field nearly an order of magnitude too small to account for the width, which is instead reproduced by a transferred $f$--$d$ superexchange of about $0.2$~K along the Gd--O--Mo bridge. The linewidth thereby measures the superexchange coupling of the probe to the orphan-spin population. Independently, the half-field forbidden satellite collapses in the magnetic host while surviving in the nonmagnetic one, a dynamical dephasing signature of the fluctuating orphan moments. The static (linewidth) and dynamic (half-field) channels together make the dilute Gd$^{3+}$ resonance a local probe of both the coupling and the dynamics of valence-bond-glass orphan spins.

\begin{figure}[t]
\centering
\includegraphics[width=\columnwidth]{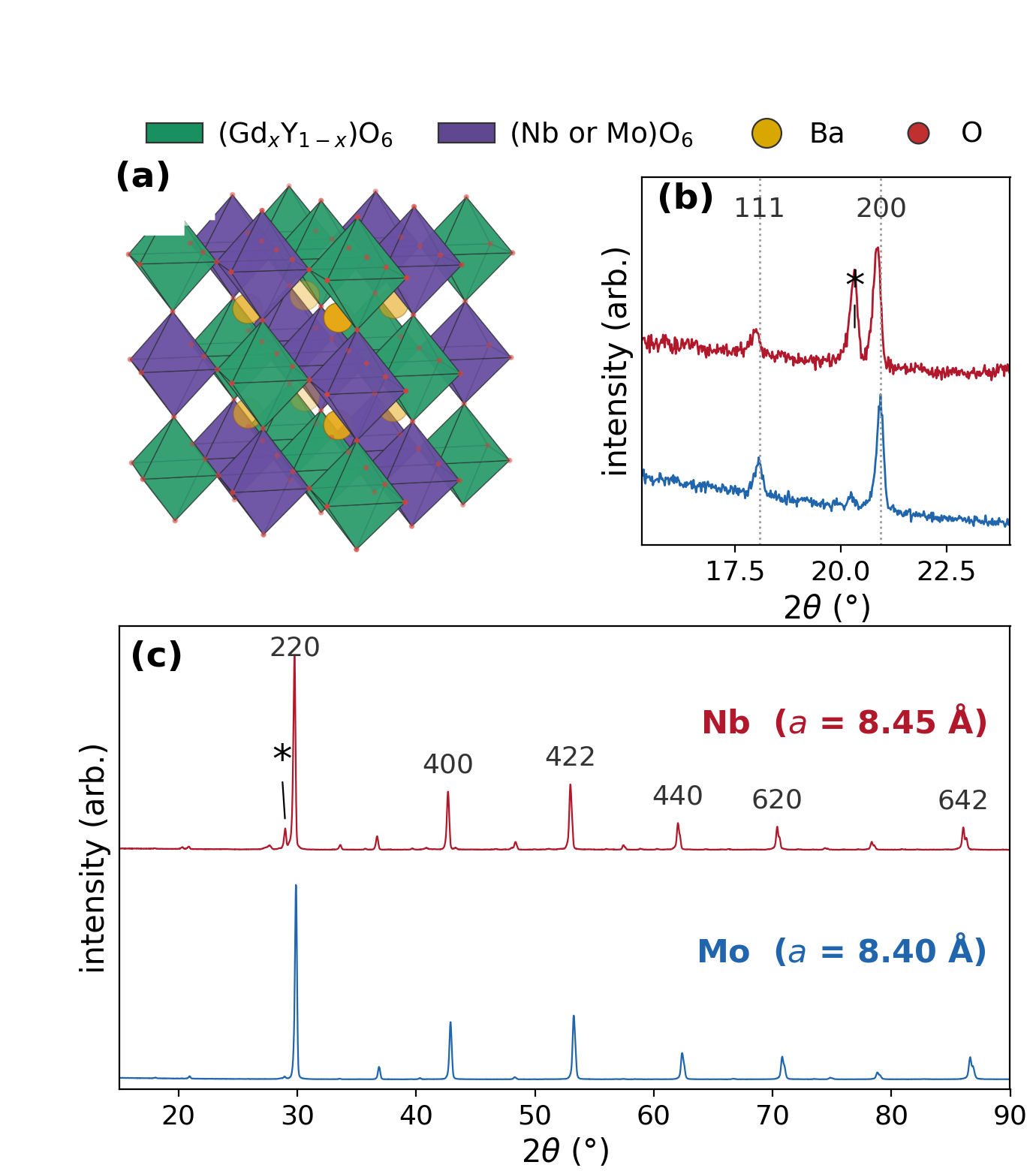}
\caption{(a) Ordered double-perovskite structure of Ba$_2$(Gd$_x$Y$_{1-x}$)B$'$O$_6$ (B$'$ $=$ Nb or Mo), with the magnetic Gd$^{3+}$ dopant dilute on the Y$^{3+}$ site. (b) Low-angle zoom of the (111) superlattice reflection that marks B-site ordering, beside the (200) fundamental reflection. (c) Powder x-ray diffraction of the $x=$ 0.05 samples (Nb, top; Mo, bottom), indexed to single-phase cubic $Fm\bar{3}m$. Asterisks in (b) and (c) mark reflections of a minor unreacted Y$_2$O$_3$ second phase, much stronger in the Nb sample; the low-angle $(211)$ line in (b) is resolved only in the niobate.}
\label{fig:structure}
\end{figure}

\section{Methods}

\subsection{Sample synthesis}
Polycrystalline Ba$_2$(Gd$_x$Y$_{1-x}$)B$'$O$_6$ (B$'$ $=$ Nb, Mo; $x = 0.02$ and $0.05$) was prepared by conventional solid-state reaction, following the procedure established for Ba$_2$YMoO$_6$~\cite{aharen2010,carlo2011}. Stoichiometric mixtures of BaCO$_3$, Y$_2$O$_3$, Gd$_2$O$_3$, and either MoO$_3$ or Nb$_2$O$_5$ were ground, fired at 950~$^\circ$C for 12~h, reground, and fired at 1250--1300~$^\circ$C, with grinding before, between, and after the sequential firings. The molybdate was fired under a reducing 5\% H$_2$/Ar atmosphere to stabilize the Mo$^{5+}$ ($4d^1$) oxidation state. The niobate (Nb$^{5+}$, $4d^0$) was prepared by the same route with Nb$_2$O$_5$ substituted for MoO$_3$ and fired in air, the $d^0$ Nb$^{5+}$ state being stable without a reducing atmosphere. Gd$_2$O$_3$ was added to the stoichiometric mixture to substitute Gd$^{3+}$ for a nominal 2\% or 5\% of the Y$^{3+}$; all compositions are quoted from the batched stoichiometry. Phase purity was verified by powder x-ray diffraction.

\subsection{Structural characterization}
Phase purity and the ordered double-perovskite structure were confirmed by powder x-ray diffraction with a Siemens D-500 diffractometer with a graphite secondary monochromator using Cu K$\alpha$ radiation ($\lambda = 1.5418$~\AA) over $2\theta = 10$--$120^\circ$ in $0.02^\circ$ steps. Patterns were indexed in the cubic $Fm\bar{3}m$ double-perovskite setting, and lattice parameters were extracted by Nelson--Riley extrapolation of the fitted peak positions. The (111) superlattice reflection, absent if the two octahedral cations are randomly mixed, confirms rock-salt ordering of the Y/Gd and Nb(Mo) sublattices. The lattice parameters of the two hosts, $a_{\mathrm{Nb}} \approx 8.45$~\AA{} and $a_{\mathrm{Mo}} \approx 8.40$~\AA, differ by about $0.6\%$, consistent with the difference between the Mo$^{5+}$ and Nb$^{5+}$ ionic radii. This establishes that the two compounds are isostructural to within our resolution, which is the control on which the linewidth comparison relies (Fig.~\ref{fig:structure}). Both hosts show weak reflections from unreacted Y$_2$O$_3$ (marked by asterisks), a minor diamagnetic secondary phase several times stronger in the niobate, consistent with the more refractory solid-state reaction of Nb$_2$O$_5$. It does not affect the dilute-Gd resonance analysis.

\subsection{Broadband CPW EPR}
Microwave transmission was measured with a vector network analyzer (VNA) through a CPW (center-conductor slot $10~\mu$m wide $\times\,100~\mu$m long) on which the polycrystalline sample was mounted. For each measurement the complex transmission $S_{21}$ was recorded as a continuous frequency sweep (2--18~GHz, 1024 points, 5 averages, 1~kHz intermediate-frequency bandwidth, 0~dBm at the probe) at each of 48 fixed magnetic fields stepped from 0.7 to 0~T. The $2\%$ molybdate linewidth was refined with a higher-precision scan of the same field range at 25 averages. This yields a two-dimensional map $S_{21}(B,f)$ (Fig.~\ref{fig:map}). The field was supplied by a superconducting magnet and the sample temperature stabilized at 5~K. The field-dependent absorption $\Delta|S_{21}|(B,f)$ was obtained by subtracting the non-resonant transmission background (Sec.~\ref{sec:analysis})~\cite{eckel2024}. Because CPW transmission mixes the real and imaginary parts of the dynamic susceptibility, each Gd$^{3+}$ resonance appears as a Dysonian (mixed absorption and dispersion) lineshape. This mixing follows the propagation and coupling phase of the transmission line, as is standard in broadband stripline and CPW magnetic resonance~\cite{kalarickal2006,bilzer2007,maksymov2015}. It is a property of the measurement geometry rather than a metallic skin effect, so the dispersion-absorption phase $\alpha$ (Sec.~\ref{sec:analysis}) is treated as an instrumental parameter and the resonance field and linewidth are the physical quantities extracted.

\subsection{Field calibration}\label{sec:fieldcal}
The magnetic-field axis was calibrated against the EPR of the free-radical standard DPPH ($g = 2.0036$)~\cite{krzystek1997}. Fine field sweeps (64 points spanning $\approx 3$~mT at $\sim 48~\mu$T resolution, 100 averages) centered on the DPPH resonance were recorded at 18 frequencies from 1 to 18~GHz. The true resonance field at each frequency was extracted by a Lorentzian fit to the phase-rotated complex $S_{21}$. The magnet controller was found to over-report the field by a small amount that grows linearly with frequency. This is consistent with a $\approx 0.35\%$ gain error in the current-to-field conversion. A linear fit to the reliable points (SNR $\geq 8$) gives the correction
\begin{equation}
B_{\mathrm{true}} = B_{\mathrm{ctrl}} - 0.15375\,f\,(\mathrm{GHz}) + 0.32496 \quad (\mathrm{mT}),
\end{equation}
with a residual uncertainty of $\pm 0.5$~mT ($1\sigma$). This correction was applied to every resonance field prior to extracting $g$. It is a property of the magnet and is independent of sample temperature.

\subsection{Lineshape analysis and uncertainty}\label{sec:analysis}
At each fixed frequency (each column of the map) the field-dependent absorption was fit to the sum of two Dysonian components (the main Gd$^{3+}$ resonance and its half-field satellite) on top of a low-order polynomial baseline representing the non-resonant CPW background. Each component,
\begin{equation}
D(B) = A\,\frac{\cos\alpha\,W^2 + \sin\alpha\,W\,(B-B_0)}{(B-B_0)^2 + W^2},
\end{equation}
is described by an amplitude $A$, a dispersion-absorption phase $\alpha$, a center field $B_0$, and a half-width $W$ (full width at half maximum $=2W$). Analysis was restricted to the resolved region $f > 8$~GHz, where the main line and its half-field satellite are cleanly separated. The $g$-factor was obtained from a through-origin linear fit of the calibrated main-line resonance field versus frequency, $f = g\,(\mu_B/h)\,B_{\mathrm{true}}$, which is insensitive to any residual constant field offset. The linewidth reported for each sample is the robust median of the per-column full widths over the resolved region.

Two sources of uncertainty are propagated. A statistical uncertainty follows from the per-column fit scatter. A background-modeling systematic was quantified by comparing the two-component fit against an independent single-component fit of the main line over a restricted field window with a local linear baseline. The two procedures agree to within $\pm 9$~mT (Nb) and $\pm 16$~mT (Mo) on the absolute linewidth, and this systematic is quoted alongside the statistical error (Fig.~\ref{fig:fwhm}). The half-field satellite intensity depends strongly on the background model and is therefore treated only qualitatively. The satellite is reported descriptively; its effective $g \approx 2\,g_{\mathrm{main}}$ identifies it as the nominally forbidden $\Delta m_S = \pm 2$ transition of the $S = 7/2$ Gd$^{3+}$ multiplet, which gains intensity through zero-field splitting and dipolar coupling and may also carry a contribution from dipolar-coupled Gd$^{3+}$ pairs. Consistent with this assignment, the satellite weakens steadily with increasing frequency, as expected for a zero-field-split forbidden transition whose intensity is borrowed through a state mixing that scales with the ratio of the zero-field splitting to the Zeeman energy and therefore falls off as the resonance field grows. No quantitative claim rests on it.

\section{Results}

Powder x-ray diffraction confirms that both compounds form the cubic ($Fm\bar{3}m$) ordered double-perovskite phase [Fig.~\ref{fig:structure}(c)]. The (111) superlattice reflection [Fig.~\ref{fig:structure}(b)] establishes rock-salt ordering of the Y/Gd and Nb(Mo) sublattices. The lattice parameters of the niobate and molybdate agree to about $0.6\%$, so the Gd$^{3+}$ dopant occupies an essentially identical crystallographic environment in the two materials. The isovalency and nearly identical ionic radii of Gd$^{3+}$ and Y$^{3+}$ make it expected that the Gd dopant substitutes on the Y site rather than the Nb/Mo site, which differs markedly in valence and radius. The $2\%$ samples show the same single-phase double perovskite with the (111) ordering reflection resolved (Supplemental Material, Fig. S1). 

%Weak reflections of unreacted Y$_2$O$_3$ (strongest the bixbyite $(222)$ line near $2\theta = 29^\circ$, marked by the asterisk in Fig.~\ref{fig:structure}(c)) appear in both hosts but are several times stronger in the niobate, consistent with the more refractory solid-state reaction of Nb$_2$O$_5$. Because Y$_2$O$_3$ is diamagnetic it gives no EPR signal and does not affect the dilute-Gd resonance analysis. 

%Because Y$_2$O$_3$ is diamagnetic and is confined to the nonmagnetic Y$^3+$ site, it cannot contribute to the linewidth contrast reported below, and it is excluded from the resonance analysis.

\begin{figure*}[t]
\centering
\includegraphics[width=\textwidth]{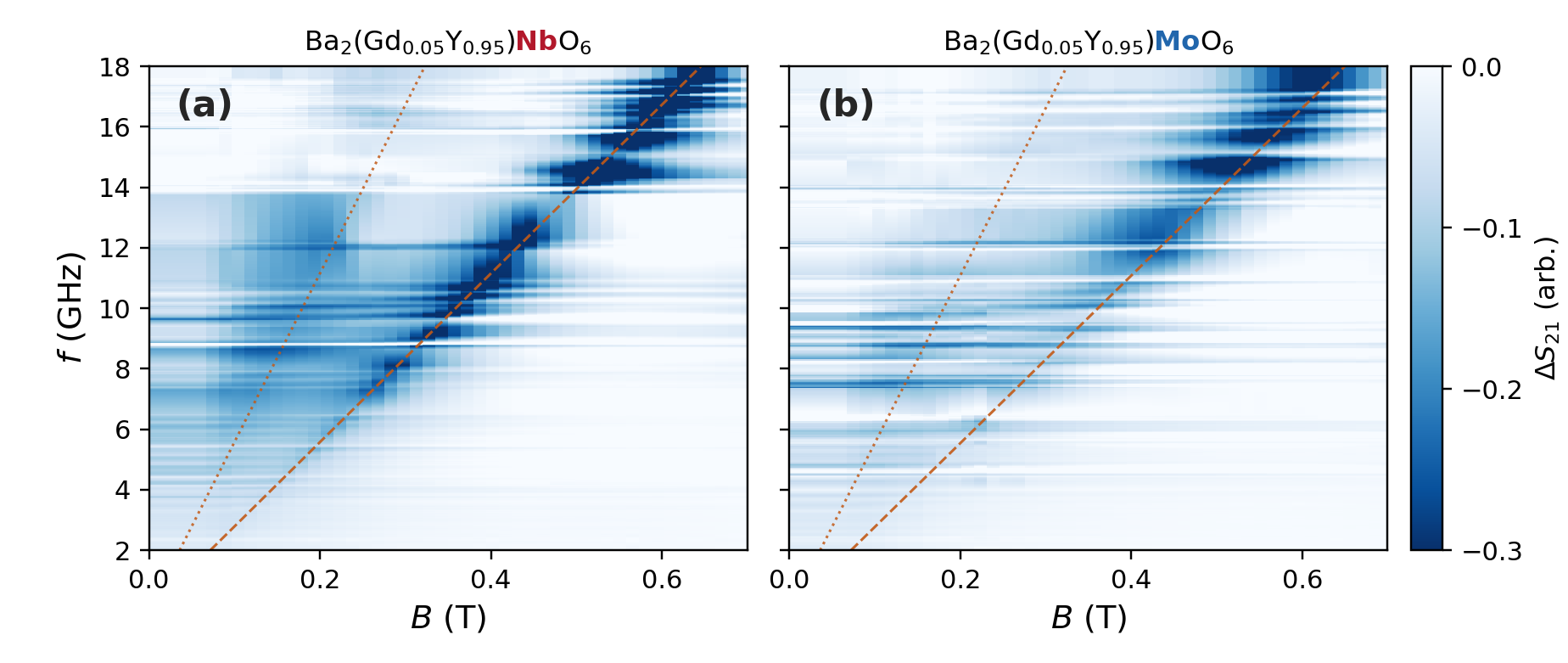}
\caption{Broadband field--frequency resonance maps at 5~K for the (a) niobate and (b) molybdate hosts at 5\% Gd; the panel titles give the full composition Ba$_2$(Gd$_{0.05}$Y$_{0.95}$)B$'$O$_6$ (B$'=$ Nb, Mo), with the B$'$-site cation highlighted. The sequential color scale is the transmission $\Delta S_{21}(B,f)$ after subtracting the non-resonant CPW background determined column-by-column in the lineshape fit (Sec.~\ref{sec:analysis}), running from white (no resonant response) to navy (deepest absorption). The main Gd$^{3+}$ resonance is the band that disperses linearly from the origin; the dashed line is $f = g\,(\mu_B/h)\,B$ with the fitted $g$ ($1.992$ for Nb, $1.979$ for Mo) from the column-by-column analysis, and the dotted line marks the corresponding half-field ($\Delta m_S = \pm 2$) satellite at $g_{\mathrm{eff}} = 2g$, which the fitted satellite centres track to within the field-step spacing. The main band disperses continuously across the full 2--18~GHz range in both hosts, and the molybdate band is visibly broader than the niobate band, consistent with the linewidth contrast quantified below. For display, frequencies coincident with cable resonances and transmission nulls are interpolated across the frequency axis, and a few narrow off-resonance instrumental features (a low-field magnet edge, a single glitched field step, and a high-field range-switch transient) are bridged by interpolation; none overlaps the Gd$^{3+}$ resonance or enters the quantitative analysis, which uses the unprocessed per-column data.}
\label{fig:map}
\end{figure*}

Figure~\ref{fig:map} presents the broadband response of both hosts as field-frequency maps of the absorption $\Delta S_{21}(B,f)$ at 5~K. In each host a single Gd$^{3+}$ resonance disperses linearly from the origin and follows the line $f = g\,(\mu_B/h)\,B$ with $g = 1.99$ continuously across the entire 2--18~GHz window. That one branch spans the full range confirms that the column-by-column scatter in $g$ above 8~GHz reflects per-column fit noise rather than multiple or shifting resonances, and the broader absorption band of the Mo host is already apparent by eye. This continuous single branch is the basis for the column-by-column lineshape analysis that follows.

\begin{figure*}[t]
\centering
\includegraphics[width=\textwidth]{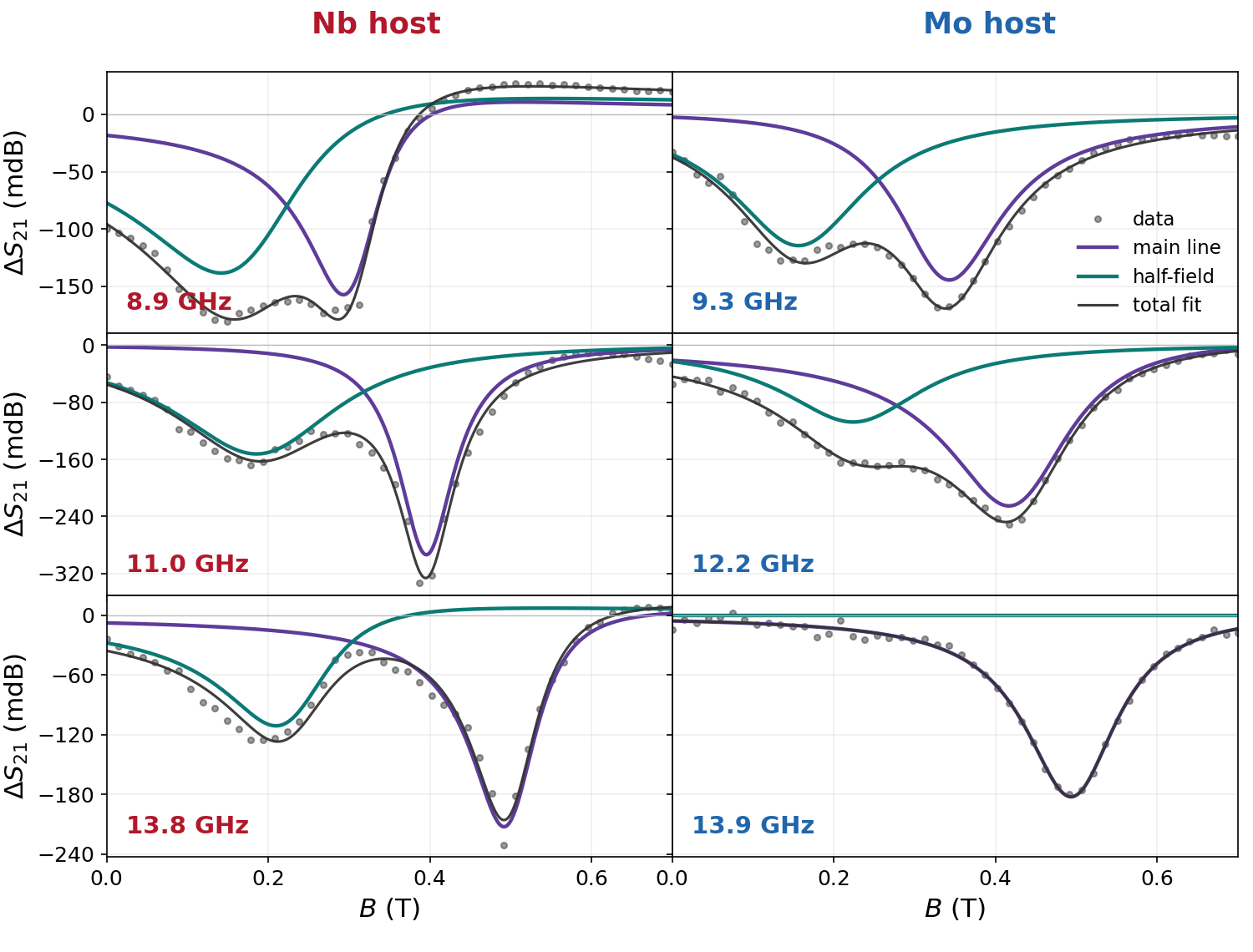}
\caption{Field-swept absorption $\Delta|S_{21}|(B)$ at 5~K for the $x=$ 0.05 Nb host (left column) and Mo host (right column) at three representative frequencies. Grey points are data. Curves show the two-Dysonian fit decomposed into the main Gd$^{3+}$ line, the half-field satellite, and their sum.}
\label{fig:fits}
\end{figure*}

Figure~\ref{fig:fits} shows representative field-swept absorption spectra $\Delta|S_{21}|(B)$ for the Nb and Mo hosts at three frequencies, together with the two-Dysonian fits. Each spectrum is well described by a main Gd$^{3+}$ resonance and a weaker half-field satellite on a smooth non-resonant background, with comparable fit quality across the resolved frequency range and for both hosts. At fixed frequency the Mo-host resonance is visibly broader than that of the Nb host.

A through-origin fit of the resonance field versus frequency gives $g = 1.992$ (Nb) and $1.979$ (Mo) at 5\% Gd. The 2\% samples give the same value, with larger per-column scatter. All four samples lie within $\pm 0.02$ of $g \approx 1.99$, just below the free-electron value ($g = 2.0036$) as expected for the $^8S_{7/2}$ Gd$^{3+}$ ion, and the per-column values are independent of frequency (Fig. S3). We therefore quote $g = 1.99(2)$, independent of host and of Gd concentration.

\begin{figure}[t]
\centering
\includegraphics[width=\columnwidth]{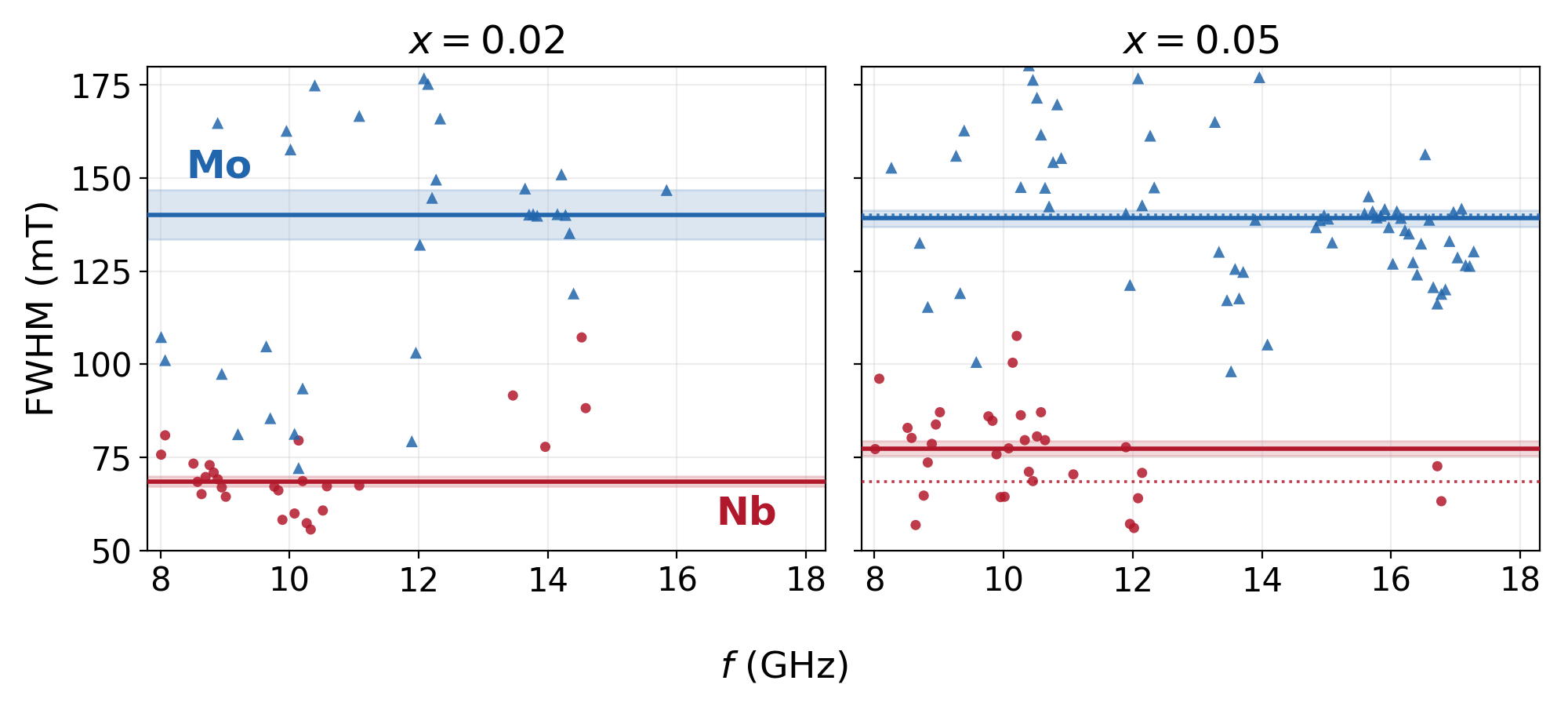}
\caption{Main-line full width at half maximum versus frequency at 5~K, for 2\% Gd (left) and 5\% Gd (right), Nb (red) and Mo (blue). Points are per-column widths; solid lines are robust-median values with $\pm 1$ standard-error bands. Dotted lines in the right panel mark the 2\% medians. From 2\% to 5\% the Nb width rises by a resolved $\approx 8$~mT (Gd--Gd dipolar broadening), while the Mo width is unchanged (concentration-independent host coupling). Because the two concentrations of a given host are analyzed by the identical procedure, the background-modeling systematic ($\pm 9$~mT Nb, $\pm 16$~mT Mo) shifts both by the same amount and so drops out of the 2\%-to-5\% difference, even though it remains in each absolute width.}
\label{fig:fwhm}
\end{figure}

The main-line linewidth is the quantity that distinguishes the two hosts (Fig.~\ref{fig:fwhm}). At 5~K the Nb-host linewidth is $76 \pm 2$~mT (statistical) and the Mo-host linewidth is $139 \pm 2$~mT, a factor of $\approx 1.8$ larger. This contrast is robust. The two background-modeling procedures (Sec.~\ref{sec:analysis}) shift the absolute widths by at most $\pm 9$~mT (Nb) and $\pm 16$~mT (Mo), and the corresponding systematic bands remain well separated. The $\approx 63$~mT gap between the hosts is therefore several times larger than any plausible analysis systematic.

The concentration dependence separates the host effect from Gd--Gd broadening (Fig.~\ref{fig:fwhm}). The Mo-host linewidth is the same between the 2\% and 5\% samples ($143 \pm 3$ and $139 \pm 2$~mT) and thus independent of Gd concentration, as expected if each probe spin samples the host magnetism independently. Measured on two independently prepared samples, the agreement also shows the width is consistent across preparations. The Nb-host linewidth, in contrast, grows with Gd content from $68 \pm 2$~mT at 2\% to $76 \pm 2$~mT at 5\%. With no B$'$-site moments the only concentration-dependent broadening is Gd--Gd dipolar coupling, which this increase isolates. Because the two Nb samples are analyzed identically, the background-modeling systematic is common to both and largely cancels in the difference. That the same Gd--Gd increment shifts the small Nb width visibly but leaves the large Mo width unchanged reflects the quadrature addition of independent broadening sources (Supplemental Material, Sec.~S5) and confirms that host coupling dominates the Mo width.

\section{Discussion}

In the nonmagnetic Nb host the only magnetic species is the dilute Gd$^{3+}$ dopant itself, so the observed linewidth reflects the intrinsic broadening of the Gd resonance. This broadening arises from unresolved $S = 7/2$ fine structure, that is, a distribution of zero-field-splitting parameters and $g$-strain from local site disorder, together with Gd--Gd dipolar coupling. The growth of the Nb-host's linewidth with Gd content (Fig.~\ref{fig:fwhm}) confirms the Gd--Gd contribution directly, with the concentration-independent single-ion terms setting the remainder. This sets the baseline linewidth against which the magnetic host is compared.

The Mo host is structurally identical, with the same $Fm\bar{3}m$ symmetry and the same Gd$^{3+}$ site; the diffraction gives lattice parameters that differ by approximately $0.6\%$ (the molybdate as the smaller), consistent with the smaller ionic radius of Mo$^{5+}$ ($0.61$~\AA) versus Nb$^{5+}$ ($0.64$~\AA)~\cite{shannon1976}. The only magnetically relevant difference is the moment on the B$'$ site, Mo$^{5+}$ ($4d^1$) in place of nonmagnetic Nb$^{5+}$ ($4d^0$). We therefore attribute the near-doubling of the Gd linewidth in the Mo host to coupling of the Gd$^{3+}$ probe spin to the Mo$^{5+}$ moments. This coupling adds a distribution of internal fields (dipolar and exchange) at the Gd site. Because the two hosts share the same Gd site and dopant concentration, the Gd--Gd dipolar broadening is identical in both and cannot account for the difference.

The excess width is magnetic rather than structural in origin. In principle a broader line could instead reflect a wider distribution of zero-field-splitting parameters or a $g$-strain at the Gd site. Three features rule this out. First, the hosts are isostructural to within $0.6\%$ and share the same $Fm\bar{3}m$ symmetry and Gd site, so the single-ion crystal-field distribution should be nearly identical and cannot plausibly double the width. Second, the width shows no systematic increase with frequency across the resolved range (Fig.~\ref{fig:fwhm}), disfavoring a $g$-strain contribution, which would grow with field, and instead pointing to a distribution of internal dipolar and exchange fields. Third, the excess does not scale with Gd concentration (Fig.~\ref{fig:fwhm}), as a Gd--Gd or dopant-disorder origin would, but is instead a per-probe coupling to the host. These features together identify the excess as a magnetic coupling of the Gd probe to the residual Mo$^{5+}$ moments, whose scale we now estimate.

Two considerations set the scale of this coupling. Each Gd$^{3+}$ probe has six nearest Mo$^{5+}$ neighbors at $a/2 \approx 4.20$~\AA{} through linear Gd--O--Mo bridges of the corner-sharing octahedra. If every Mo carried its full paramagnetic moment ($\mu_{\mathrm{eff}} \approx 1.4~\mu_B$)~\cite{aharen2010}, the Van Vleck dipolar second moment would give a local-field spread of $\approx 23$~mT (root mean square), or $\approx 50$~mT full width (Supplemental Material, Sec.~S4). At 5~K, however, most Mo moments are locked into non-magnetic singlets that fluctuate at the spin-gap scale, far faster than the EPR timescale, and are averaged out. Only the residual ``orphan'' spins, a fraction $\approx 8\%$ of the Mo moments between 1.9 and 7~K~\cite{devries2013}, stay paramagnetic and quasi-static (fluctuating on a $\sim$5~ns timescale~\cite{devries2013}) as seen by the probe. For this dilute set the dipolar second moment scales with the fraction, so the root-mean-square spread falls to $\sqrt{0.08}\times 23 \approx 6$~mT ($\approx 15$~mT full width). Because independent broadenings add in quadrature, the host contribution to the Mo linewidth is $\sqrt{139^2-76^2} \approx 116$~mT (Supplemental Material, Sec.~S5), and this orphan dipolar term supplies less than an eighth of it.

That dipolar coupling falls short is the crux of the mechanism. With the singlet majority magnetically silent, the sparse orphan spins alone would give only $\approx 15$~mT of broadening, nearly an order of magnitude below the host contribution. The missing broadening is supplied by a transferred $f$--$d$ superexchange along the same Gd--O--Mo bridge (Gd $4f$--$5d$--O $2p$--Mo $4d$)~\cite{anderson1959,goodenough1955,kanamori1959}, which couples far more strongly per neighbor than the dipolar field: an exchange constant of only $J \approx 0.2$~K acting on the orphan spins reproduces the measured width (Supplemental Material, Sec.~S4). Such a scale is typical for Gd superexchange in oxides, consistent with the generally weak, sub-kelvin coupling of the shielded $4f$ shell---for example the $\approx 0.1$~K nearest-neighbor exchange in Gd$_3$Ga$_5$O$_{12}$~\cite{schiffer1994}. Because Gd occupies the Y site, its bridge to Mo shares the oxygen that carries the Mo--O--Y--O--Mo intersublattice path. That a core-like $4f$ ion registers a measurable superexchange at all reflects the same Mo $4d$--O $2p$ covalency that underlies the strong Mo--Mo superexchange of the valence-bond glass, even though the $f$--$d$ and $d$--$d$ couplings differ in both magnitude and path. This is a striking outcome for a nominally weakly coupled $S$-state probe: the Gd linewidth is set not by the magnetic dipole fields of its neighbors but by superexchange to the residual orphan moments, so the width becomes a direct, local measure of the orphan-spin population and its coupling in the valence-bond glass. 

%Because only a minority of Gd sites border an orphan, a mild core-plus-wing lineshape is expected; a quantitative lineshape decomposition is left to the temperature- and concentration-dependent study noted below. All of these effects are magnetic in origin and absent in the nonmagnetic Nb host.

The $g$-factor, in contrast, is the same in both hosts within uncertainty. Because $g$ is fixed by the mean resonance field while the linewidth reflects the spread of fields over the Gd sites, the Mo host broadens the line without moving its center. This is the signature of a random, zero-mean distribution of internal fields, in which the disordered Mo moments shift each Gd site differently and widen the resonance while leaving the average field, and hence $g$, unchanged. A coherent internal field, identical at every site, would instead have displaced the resonance and shifted $g$, and its absence reinforces the disordered, glassy picture that the linewidth already indicates. Any residual shift, from a slight field-induced polarization of the Mo moments, would be proportional to the host susceptibility, which is small for the gapped valence-bond glass at 5~K and below our $\pm 0.02$ resolution.

The half-field satellite opens a second, dynamical channel onto the same orphan spins. Being a forbidden $\Delta m_S = \pm 2$ transition, its intensity is borrowed coherently from the zero-field-splitting admixture and is therefore more sensitive to dephasing than the allowed main line. In the nonmagnetic Nb host the satellite stays resolvable across the full $8$--$18$~GHz band, whereas in the Mo host it is lost above roughly $10$--$12$~GHz. The contrast is already visible in Fig.~\ref{fig:fits} at nearly the same frequency, where the satellite is clearly present in the $13.8$~GHz Nb cut but has vanished in the $13.9$~GHz Mo cut, even though its peak height is comparable in the two hosts where both are resolved (Supplemental Material, Fig. S4). This collapse is far larger than the modest broadening the host adds to the satellite width, so it reflects a \emph{quenching} of the forbidden transition rather than a broadening of it: the fluctuating orphan Mo moments dephase the coherent zero-field-splitting admixture on which the satellite depends. Where the linewidth measures the static superexchange coupling to the orphan spins, the half-field collapse is sensitive to their fluctuation dynamics, so the two observables probe complementary---static and dynamic---aspects of the same orphan-spin population.

\section{Conclusion}

Electron paramagnetic resonance of a dilute Gd$^{3+}$ probe, measured by broadband coplanar-waveguide transmission, provides a local spectroscopy of the orphan spins of the valence-bond glass Ba$_2$YMoO$_6$, with the isostructural nonmagnetic niobate as an equivalent-lattice control. The Gd$^{3+}$ linewidth nearly doubles in the magnetic host, but the sparse orphan spins are far too dilute for dipolar coupling to account for it; the excess instead reflects a transferred $f$--$d$ superexchange of $J \approx 0.2$~K along the Gd--O--Mo bridge, so the exchange-dominated linewidth is a static, local measure of the orphan-spin coupling. The half-field forbidden satellite collapses in the magnetic host while persisting in the nonmagnetic one, a dynamical dephasing signature of the same fluctuating moments, and the host-independent $g$-factor identifies the internal field as a zero-mean random distribution. The static (linewidth) and dynamic (half-field) channels thus probe complementary aspects of the same orphan-spin population at a single crystallographic site. Because the probe couples to the Mo$^{5+}$ moments through a superexchange bridge rather than through weak dipolar fields, the dilute Gd$^{3+}$ resonance offers a sensitive, site-resolved window on the valence-bond-glass state of the Ba$_2$YMoO$_6$ family, and more broadly on the residual moments of quantum-disordered insulators.

\begin{acknowledgments}
We thank Bryan Eigenbrodt (Department of Chemistry, Villanova University) for assistance with sample synthesis, and Tommaso Cappello (Department of Electrical and Computer Engineering, Villanova University) for assistance with CPW fabrication on his LPKF S64 circuit-board milling machine. This work was supported by the National Science Foundation under CAREER Award No.~DMR-1943389 (S.D.).

\end{acknowledgments}

% ---------------- References ----------------

\end{document}